\documentclass[useAMS,usenatbib,usegraphicx]{mn2e}
\begin{document}

\title [Models of NGC~2207 and IC~2163]
{The grazing encounter between IC~2163 and NGC~2207: 
Pushing the limits of observational modeling 
$\thanks{Based in part on observations with the NASA/ESA
Hubble Space Telescope, obtained at the Space Telescope Science
Institute, which is operated by the Association of Universities for
Research in Astronomy, Inc. under NASA contract NAS5-26555}$ 
$\thanks{E-mail: curt@iastate.edu (CS); bge@watson.ibm.com (BGE);
elmegreen@vassar.edu (DME); rallis@mps.ohio--state.edu (MK);
ebrinks@star.herts.ac.uk (EB); magnus@oso.chalmers.se (MT) }$}  

\author [C. Struck et al.]
{Curtis ~Struck,$^1$ Michele ~Kaufman,$^2$  Elias ~Brinks,$^3$
Magnus ~Thomasson,$^4$
\newauthor Bruce G. ~Elmegreen,$^5$ 
Debra Meloy ~Elmegreen$^6$\\
$^1$Department of Physics and Astronomy, Iowa State University, Ames,
IA 50011\\
$^2$Department of Physics and Department of Astronomy,
Ohio State University, 174 W. 18th Avenue, Columbus, OH 43210\\
$^3$ Instituto Nacional de Astrof\'{\i}sica, \'Optica y Electr\'onica,
Apdo. Postal 51 \& 216, Puebla, Pue 72000, M\'exico, current
address:\\
University of Hertfordshire, School of Physics, Astronomy and
Mathematics, College Lane, Hatfield  AL10 9AB, England\\
$^4$Onsala Space Observatory, S--439 92 Onsala, Sweden\\
$^5$IBM Research Division, T.J. Watson Research Center,
P.O. Box 218, Yorktown Heights, NY 10598\\
$^6$Department of Physics and Astronomy, Vassar College,
Poughkeepsie, NY 12604 \\}

\date{Accepted .........;
      Received ..........;
      in original form ........}

\pagerange{\pageref{firstpage}--\pageref{lastpage}}
\pubyear{2005}

\maketitle

\label{firstpage}

\begin{abstract}

We present numerical hydrodynamical models of the collision between
the galaxies IC~2163 and NGC~2207. These models extend the results of
earlier work where the galaxy discs were modeled one at a time. We
confirm the general result that the collision is primarily planar,
that is, at moderate inclination relative to the two discs, and
prograde for IC~2163, but retrograde for NGC~2207. We list 34 specific
morphological or kinematic features on a variety of scales, found with
multi-waveband observations, which we use to constrain the models. The
models are able to reproduce most of these features, with a relative
orbit in which the companion (IC~2163) disc first side-swipes the
primary (NGC~2207) disc on the west side, then moves around the edge
of the primary disc to the north and to its current position on the
east side. The models also provide evidence that the dark matter halo
of NGC~2207 has only moderate extent. For IC~2163, the prolonged
prograde disturbance in the model produces a tidal tail, and an oval
or ocular waveform very much like the observed ones, including some
fine structure. The retrograde disturbance in the model produces no
strong waveforms within the primary galaxy. This suggests that the
prominent spiral waves in NGC~2207 were present before the collision,
and models with waves imposed in the initial conditions confirm that
they would not be disrupted by the collision. With an initial central
hole in the gas disc of the primary, and imposed spirals, the model
also reproduces the broad ring seen in HI observations. Model gas disc
kinematics compare well to the observed (HI) kinematics, providing
further confirmation of its validity. An algorithm for feedback
heating from young stars is included, and the feedback models suggest
the occurence of a moderate starburst in IC~2163 about 250 Myr ago.

% and minimal compression and induced star formation in the core of
% NGC~2207.

We believe that this is now one of the best modeled systems of
colliding galaxies, though the model could still be improved by
including full disc self-gravity. The confrontation between
observations and models of so many individual features provides one of
the strongest tests of collision theory. The success of the models
affirms this theory, but the effort required to achieve this, and the
sensitivity of models to initial conditions, suggests that it will be
difficult to model specific structures on scales smaller than about a
kiloparsec in any collisional system.
\end{abstract}

\begin{keywords}
galaxies: individual: (NGC 2207, IC 2163) --galaxies: interactions --
galaxies: dynamics.
\end{keywords}

\section{Introduction} 

Collisions and interactions between galaxies in groups almost
inevitably lead to mergers, with orbital energy converted into random
motions of the constituents, and into the kinetic energy of tidal tail
ejecta.  However, this process is not direct; the initial and final
relaxed states are separated by a period in which energy is also used
to generate and maintain large-scale coherent structures.

Unless the collision involves special symmetries, the complexity of
the tidal structures (and the diversity of galaxy morphology)
increases for a time of order several dynamical times (e.g., review by
\citealt{str99}). Ultimately, phase mixing and thermalization erase
these structures.  The form of the waves generated before this erasure
is a function of both the precollision morphology of the individual
galaxies and the collision parameters. Dynamical modeling can be used
to derive the initial conditions and collision parameters from
observations of the tidal structure, a kind of galaxy tomography.

One obvious difficulty in such a program is that we only see colliding
galaxies in projection in both coordinate and velocity space. However,
we can hope to compensate for the loss of 50\% of phase space
coordinate information with increasingly detailed, high resolution
observations and similarly high resolution models. New instrument and
computer technology are providing rapid improvements in both models
and observations. On the observational side, HST observations of
collisional systems like NGC~2207/IC~2163 (see Figure 1)
provide a prime example of the results of improved resolution.

\begin{figure*}
%\scalebox{0.4}{ \includegraphics{fig1.eps}}
% \vspace{20.cm}
%\includegraphics{f1.ps}
\caption{Grey-scale HST B-band image of NGC~2207 (right) and IC~2163
(left), from \citet{elm01}. Rectangles and labels show peculiar
emission features described in that paper.}
\label{fig1}
\end{figure*}

Another fundamental problem is that colliding galaxies have a wide
range of dynamical timescales. Orbital and epicyclic periods may range
from $3 \times 10^6$ yr in the central regions to more than $10^9$
yr for the orbits of stars and gas clouds flung out in long tidal
tails.  Similarly, the duration of the strong interaction in a first
encounter may be short compared to the time between encounters.

A third difficulty is that the timescale of the tidal driving is
comparable to that of the internal dynamical timescales.  For example,
waves in the inner-to-middle discs from a first disturbance may have
begun to phase mix at the onset of a renewed disturbance, which will
generate a new set of waves. The multiple sets of waves generated by
multiple impulsive disturbances will interfere with each other in ways
that depend on the degree to which the waves are material or phase
phenomena. Secondary disturbances result from time-delayed effects
like the slow fall-back of material pulled out in long tidal
structures, and the different response times of different galaxy
components (bulge, disc, and halo).

These difficulties make it nearly impossible to reconstruct the
collisional history of systems in an advanced state of merger.
Reconstruction is much more feasible in recent collisions (i.e.,
``young'' systems in a relative sense), or collisions with special
symmetries, where there are additional constraints on the collision
parameters.  Two special cases with both these attributes are the
collisional ring galaxies (see review by \citealt{app96}, and e.g.,
\citealt{str03}, \citealt{hea01}), and the ocular or eye-shaped
galaxies, like IC~2163 (see Fig. 1).

In the case of the ocular galaxies, models show that the ocular
caustic wave persists for only a relatively short time, and the
caustic is only produced in near-planar, prograde encounters, as seen
from the ocular (see \citealt{elm91}, \citealt{sun89},
\citealt{don91}). We use the term ``caustic'' here to denote a region
of stellar orbital overlap or crossing (\citealt{str90}).  The
structure of the ocular caustic also depends on the overall matter
distribution of the galaxy containing it. Thus, the presence and shape
of the ocular put strong constraints on the orbital parameters and the
global structure of the ocular galaxy.

In the NGC~2207/IC~2163 system there are additional constraints. The
two galaxies appear to be very close, at least in projection (see
Fig. 1), and have the same (l.o.s.) systemic velocity, within the
uncertainties (see Table 1 and Elmegreen et al. 1995a
\citep{elm95a}). We will show below that there are reasons to believe
that IC~2163 is physically close to the disc of NGC~2207. The HI
kinematics presented in \citealt{elm95a} also suggest that we are
looking at NGC~2207 somewhat close to face-on, so the plane of the sky
nearly coincides with the disc plane.  With this information, and the
fact that the ocular caustic takes some time to develop, we conclude
that the relative orbit is probably well bound at present.  If it were
not, and the high relative velocities were confined to the plane of
the sky (since no high velocities are found in the observations, see
\citealt{elm95a}), then the two galaxies would not stay near the point
of closest approach for long, but quickly move apart.  The relative
orbit of IC~2163 must have kept it close to NGC~2207 for the time
needed to form the ocular.  Thus, in this case we can also constrain
orbital velocity components which are not normally observable.

These additional constraints lead to the expectation that we may be
able to model this system with unusual accuracy. A number of these
constraints were already used in the earlier papers of this series to
produce preliminary models (see Elmegreen et al. 1995b
\citep{elm95b} and Elmegreen et al. 2000a \citep{elm00a}). In the
previous encounter models, the response of only one component (stars
or gas) of one galaxy was modeled and the other galaxy was represented
by a softened point mass. In the new models the responses of stars and
gas particles in both galaxies are modelled in three dimensions. This
provides a more realistic treatment of the encounter, and explains
many observed features not accounted for by the older models.

In fact, the wealth of spatial structure in this system, observed in a
variety of wavebands, presents a modeling challenge. The individual
structures, seen on a wide range of scales, depend on the
distributions of matter within the galaxies (at present and in the
recent past), and the collision parameters.  A model that was
successful in fitting most or all of these features would yield much
information about these quantities. Thus, we can hope to achieve more
than the usual modeling goal of accounting for the observed properties
of a few major tidal features.

We will see below that the best models do indeed succeed in
reproducing almost all of the major features of this system, and
provide us with a much improved understanding of it.  We begin in the
next section with a survey of the major structures observed in this
system.

\section{Overview of the NGC~2207/IC~2163 System} 

This system is relatively nearby (the mean recession velocity is about
$2750\ $km s$^{-1}$, see Table 1 and \citealt{elm95a}), so we are able
to resolve a great deal of detailed structure within it using HST. The
major structural features, as well as some finer structures, are
listed in Tables 1 and 2.  We briefly review them in this section,
emphasizing their role in constraining the models.  Details on the
observations can be found in \citet{elm95a}, \citet{elm00a}, and
Elmegreen et al. (2001 \citep{elm01}).

\begin{table*}
 \centering
 \begin{minipage}{200mm}
\caption{Observed Properties That Constrain the Models: I. Large Scale
Structures.}   
  \begin{tabular}{@{}cl@{}} 
A. Both Galaxies
\footnote{Properties derived from data presented in \citet{elm95a},
\citet{elm00a}, and \citet{elm01}} &\\[10pt] 
 1. & Distance = 35 Mpc. \footnote{1.0 arcsec = 170 pc, for 
     $H_o = 75\ $ km s$^{-1}\ $Mpc$^{-1}$}\\
 2. & Luminosity ratio (IC~2163/NGC~2207): 0.6 in NIR and 0.3 in the 
B-band.\\
 3. & Optical, isophotal major radius ($R_{25}$): 
91'' (IC~2163) and 130'' (NGC~2207).
     \footnote{Note: the IC~2163 value includes part of the tail.}\\
 4. & V$_{sys} = 2765 \pm 20\  $km s$^{-1}$ for IC~2163 and $2745 \pm
      15\  $km s$^{-1}$ for NGC~2207.
     \footnote{Heliocentric, optical definition.}\\
 5. & Side nearest us: N by NW for IC~2163, and between N and NE for 
       NGC~2207.\\
 6. & Separation between centers: 85''.
     \footnote{The western half of IC~2163 is partially obscured by
NGC~2207.}\\ 
 7. & Global SFR/(HI mass) - typical of noninteracting spirals.\\
 8. & Widespread, high velocity dispersion ($30-50\  $km s$^{-1}$) HI gas.\\
 9. & Molecular gas (traced by CO) in both discs, with more in IC
2163 and no\\
  & concentration on massive HI clouds (see 28 in Table 2).
  \footnote{See \citet{tho04}.}\\
\\
B. IC~2163 &\\
10. & Central eye-shaped oval with projected major axis length = 43''.\\ 
11. & Enhanced emission from ocular rim. 
	\footnote{This includes emission in radio continuum, HI,
optical and the near infrared.}\\
12. & Position angle of the major axis:\\
&a) Photometric PA of eye-shaped oval: $128^{\circ} \pm 3^{\circ}$.\\
&b) HI kinematic $65^{\circ} \pm 10^{\circ}$.\\
13. & Two tidal arms located symmetrically on opposite sides of the
nucleus.
	\footnote{In the sky plane, bright HI emission from the tidal
tail extends to a distance of 110'' NE from the nucleus, which is
about 2.5 times\\
 the major axis of the eye-shaped oval. Faint HI
emission from the tail extends to {\it R} = 190''. Optically, the tidal tail
extends out to at\\
least {\it R} = 90''.}\\
14. & $100\   $km s$^{-1}$ streaming motions on tidal tail and around the
eye-shaped oval.\\
15. & Mean velocity is nearly constant along the outer extension of
the tidal tail.\\
16. & Stellar arm contrast of the tidal tail is large compared to
normal spiral arms.\\
17. & The nearly-circular shape and orientation of the spiral arms
 inside the eye-shaped oval.
	\footnote{Their structure suggests that they may have been
produced by inner Lindblad resonances in the tidal potential that
formed the oval\\ 
(see \citealt{elm00a}).}\\
 
18. & Long parallel dust filaments on the tidal tail.
	\footnote{These may have originated as normal flocculent
spiral arms present before the encounter, which were stretched into
parallel filaments\\
 as the tidal tail formed.}\\
19. & Brightest CO emission from the centre position, also bright
emission\\ 
  & from oval rim.
	\footnote{The resolution is 43'' (HPBW), so emission centre is
uncertain. Spectrum is double-peaked, hinting that some emission is
from the\\
 oval rim rather than the nucleus.}\\
\\
C. NGC~2207\\
20. & Two long grand-design spirals dominate much of the disc.\\
21. & HI in the main disc forms a broad, clumpy ring, broken in the
S.\\ 
22. & A large, elliptically-shaped HI feature extends 220'' S/SE of
the nucleus.
	\footnote{This feature also contains optical filaments.}\\
23. & Position angle of the major axis: \\
&a) Photometric PA (optical and HI) of main disc:  $110^{\circ} \pm
7^{\circ}$.\\ 
&b) HI kinematic PA: varies over range $150^{\circ} - 177^{\circ}$ as
{\it R} increases.\\
&c) Isophotal on scale of S/SE extended HI pool: $160^{\circ} \pm
5^{\circ}$.\\ 
24. & Global S-shaped distortion of the HI velocity field.
	\footnote{This plus the large misalignment between photometric
and kinetic major axes imply that the disc is warped, and
there are\\
 large amplitude z-motions producing the warp.}\\
25. & Radio continuum emission is enhanced on the companion side along
the outer\\
& \ \  part of the disc.\\
26. & CO emission not concentrated on the nucleus. Mainly found
about halfway\\
 & between the centre and HI ring, at least on western half of the
disc.\\ 
 & Brightest emission at about 24'' NW of the nucleus.\\

\end{tabular}
\end{minipage}
\end{table*}

\begin{table*}
 \centering
 \begin{minipage}{200mm}
\caption{Observed Properties That Constrain the Models: II. Small Scale
Structures.}
  \begin{tabular}{@{}cl@{}}
\\
A. Both Galaxies
\footnote{Properties derived from data presented in \citet{elm95a},
\citet{elm00a}, and \citet{elm01}. Numbering continued from Table 1.}
&\\[10pt]
27. & SF regions have a normal luminosity-size distribution and
luminosity function,\\
& except with the SSC distribution weighted towards more massive
clusters.\\ 
28. & Massive ($10^8 M_{\odot}$) HI clouds in the disc, not associated
with major stellar clumps.\\ 
29. & Widely distributed young star clusters and super star
clusters. 
	\footnote{The SSCs seem to be confined to a few segments of
the spiral arms of NGC~2207 or to the rim of the eye-shaped oval in 
IC~2163.}\\ 
\\
B. IC~2163&\\
30. & The nucleus is not detected in radio continuum to a rather faint
level.\\ 
\\
C. NGC~2207&\\
31. & Chaotic network of dust spirals in nuclear region.\\
32. & Parallel, knotty dust filaments span full width of spiral arm
that is backlit by IC~2163.\\
33. & There is an unusually luminous SF region on the outer western
arm (Region i, on the \\
&anti-companion side).
	\footnote{This region is very bright in the radio continuum
and in H${\alpha}$. It coincides with a very dark dust cone, and is
detected in\\
 CO J = 1-0 emission.}\\
34. & The nucleus is not the brightest radio continuum source in the
galaxy.\\ 

\end{tabular}
\end{minipage}
\end{table*}

First of all, we note that although NGC~2207 has larger optical and HI
diameters than IC~2163, it is not proportionately more
luminous. Estimates of the luminosity ratio suggest that the masses of
the two are not greatly different, assuming that their mass-to-light
ratios are comparable, though the luminosity of IC~2163 is somewhat
less than that of NGC~2207.

\citet{elm95a} and \citet{elm95b} discussed the orientations of the
optical discs, based on the HI line-of-sight velocity fields and the
optical images.  Because of the tidal distortions, accurate values of
the orientation parameters could not be obtained by using only the
apparent axis ratio.  Presently, IC~2163 has an intrinsically oval
disc, which results in a $63^{\circ}$ offset between its photometric
and kinematic major axes. \citet{elm95a} and \citet{elm95b} chose the
projection line-of-nodes of IC~2163 (the intersection of the plane of
the disc with the plane of the sky) to be the kinematic major axis,
position angle $= 65^{\circ}$, and estimated the inclination {\it i}
(where {\it i} = 0 for face-on) as $40^{\circ}$. They interpreted the
S-shaped velocity field of NGC~2207 and the $40^{\circ}$ offset
between the photometric and kinematic major axes of its central disc
in terms of a warp that is increasing with time; this gave an
inclination for the central disc of about $35^{\circ}$ with the
projection line-of-nodes at PA $= 140^{\circ}$. These values derived
from preliminary models formed a starting point for our study. In our
simulations, we take the initial discs of these galaxies as unwarped.

As noted above, the presence of an ocular morphology in IC~2163
provides a strong constraint on the encounter dynamics.  Similarly,
both the global structure of the tidal tail (e.g., length and width),
as well as its detailed morphology, constrain both the relative orbit
and the halo structure of IC~2163. These features will be the focus of
much discussion below.

The spectacular visual impression made by this system owes much to the
spiral arms of NGC~2207, seen as emission structures over the bulk of
the galaxy, but also as backlit absorption features in front of
IC~2163. Although in the former case they are not entirely regular in
appearance, we believe that they are generally well accounted for as a
simple two-armed ({\it m=2}) density wave.  As discussed below (also
see \citealt{elm95b}), they would not be produced in otherwise
plausible models in which IC~2163 orbits in a retrograde sense
relative to NGC~2207.  (The absence of prominent tidal tails or a
bridge from NGC~2207 also supports the idea that the encounter is
retrograde for it.)  Thus, we believe that the {\it m=2} density wave
predates the current encounter, and we can test its persistence
through the encounter. This is an unusual experiment in the field of
colliding galaxy models, though the superposition of induced and
pre-existing waves is probably a common occurence in nature.

\begin{figure*}
% \vspace{20.cm}
%\includegraphics{f2.ps}
\caption{Contours of the total HI emission, which is kinematically
associated with NGC~2207, on a grey scale image of the system from the
digital sky survey.  The line-of-sight column density contour levels
are 6.4, 13, 19, 25, 32, 38 and 44 atoms cm$^{-2}$, and the massive HI
clouds in NGC~2207 are labeled.}
\label{fig2}
\end{figure*}

\begin{figure*}
% \vspace{20.cm}
%\includegraphics{f3.eps}
\caption{Contours of the total HI emission kinematically associated
with IC~2163. The HI contour levels and the grey scale image are the
same as in Fig. 2. Labels identify massive HI clouds in
IC~2163. Region i, marked here and in Fig. 1, is unusually luminous in
H$\alpha$ and in the radio continuum. It is not associated with a
massive HI cloud (cf. Fig.2).}
\label{fig3}
\end{figure*}

The large-scale distribution of HI in NGC~2207 differs both in shape,
and by its significant southern extension, from the global stellar
distribution (see Figures 2 and 3). Although faint optical filaments
are visible in the southern extension (see \citealt{elm00a}), the
large southern extension is predominantly an HI feature. In the main
disc of NGC~2207 the HI distribution forms a broad clumpy ring, broken
in the south. The ring contains massive ($10^8\ M_{\odot}$) clouds,
which do not always coincide with the stellar arms. The ring-like
distribution may be largely a result of a gas hole in the central
regions. This hole is not entirely filled by molecular gas either,
since the molecules in NGC~2207 are concentrated on the western side
of the disc (see Table 1). The gas hole in the central region may be
the result of the weak central bar as noted in Paper I. We have not
attempted to model the bar. The models also suggest that a mild
collisional ring ({\it m=0}) component to the collisional perturbation
may play a role in generating the HI ring.

Both the optical \citep{elm00a} and HI observations \citep{elm95a}
suggest the presence of a tidal bridge from IC~2163 to
NGC~2207. (E.g., see Fig. 3, where the cloud labeled I4 is on the
tidal bridge.) We use the models below to argue that development of
the bridge may be related to the mysterious radio continuum ridge
found in the NE quadrant of NGC~2207, lying nearly on a tangent to
NGC~2207, and connecting to IC~2163.

Region i (see Fig. 1) on the outer western arm of NGC~2207 is also
unusually luminous in the radio continuum and H$\alpha$. It contains a
dark dust cone and arcs of star clusters, and is associated with CO
1-0 emission.  This is one of the most mysterious features in the
system, though the modelling below does suggest some possible
explanations. In general, it is also remarkable that most of the radio
continuum emission comes from the outer parts of the NGC~2207 disc,
rather than the nuclei of the two galaxies.

\citet{elm95a} displays the kinematical information from a VLA HI cube
as channel maps, velocity distributions, velocity fields, and velocity
dispersion maps. In that paper, the HI contributions of the two
galaxies were separated by inspection of the channel maps in a
movie-mode. At some locations with overlap in projection, the
kinematic distinction between the gas in each was clear. If not, then
in overlap areas, 80\% of the gas was attributed to NGC~2207, the same
as the global HI ratio. That paper interprets these data in terms of
streaming motions, widespread high velocity dispersions ($30-50$ km
s$^{-1}$) in the HI gas, warping and z-motions. There are 11 massive
($10^8\ M_{\odot}$) clouds in the discs and tidal arms of these
galaxies. For comparison below with the kinematic results of our new
models, we have constructed summed position-velocity diagrams from
this HI cube (see Figure 9).

Table 2 contains a list of relatively small-scale features.  First,
the star-forming properties of this system are not spectacular.  The
global SFR is modest, and the young cluster luminosity functions of
both galaxies do not differ significantly from those of typical
late-type disc galaxies (\citet{elm01}).  A relatively small number of
super star clusters (SSCs), like those in other interacting systems,
has been found.  Interestingly, these are found in the outer regions
of the two discs, rather than concentrated in the central or nuclear
regions.

Besides the star clusters, there are also a number of other peculiar
emission features found asymmetrically distributed in the outer parts
of NGC~2207 (mostly in the southern and western half of the disc, see
Fig. 1).  Collectively, the set of these emission features is unique
to the best of our knowledge (see Sec. 7 in \citealt{elm00a}).
However, this may be partly the result of the superb HST resolution of
this nearby system.  Alternately, they may involve either mass
transfer from an earlier encounter, or continuous interaction between
the two discs for some time before the present.  Either of these
alternatives may help understand the concentration of peculiar
features opposite the current locus of interaction.

Other features include the numerous, long, thin, and highly contrasted
filaments seen in both the tidal tail of IC~2163, and in the NGC~2207
spiral segments projected onto the face of IC~2163.  They also include
the massive HI clouds (Figs. 2 and 3) and the inner spirals and
turbulent appearance of the nuclear regions of both galaxies.

\section{Model Description}

Previous models of this system captured a number of important
features, but were more approximate than our new models.
\citet{elm95b} used an N-body two-dimensional (modeling the primary
disc plane) simulation of the effect of the encounter on the stellar
disc of IC~2163 with NGC~2207 represented by a point mass; the gas
disc was not treated.  This paper also presented a two-dimensional
(azimuthal $\theta$, z) model for NGC~2207 in which the disc of
NGC~2207 was divided into segments, no radial motions were allowed,
and IC~2163 was represented by a point mass. \citet{elm00a} presented
a three-dimensional SPH simulation of the gaseous disc of IC~2163, not
the stellar disc, with NGC~2207 represented by a softened point mass.

The new models in this paper use two-disc, gas-star, three-dimensional
SPH codes to model the discs of both galaxies simultaneously
throughout the encounter. We can thus see if the present more
sophisticated models reproduce the same types of unusual features as
the more approximate early models, and if the new models can explain
observed features not accounted for by the older models.  For this
purpose, we carried out several dozen new simulations with two
different gas-star SPH codes to model the detailed morphology and
kinematics of the system. We present the most successful of these
models in the following sections. In this section we briefly describe
the simulation codes, scale factors, and initial conditions used in
these models.

\subsection{The S97 Code}

The S97 code, used here and in \citealt{elm00a} (and described in
detail in \citealt{str97}) is designed for three-dimensional
simulations of gas and star particles in rigid halo gravitational
potentials. This code uses an SPH algorithm for the gas dynamics, with
hydrodynamic forces computed with a spline kernel on a grid. The grid
has a fixed unit cell size, but varying extent. A standard artificial
viscosity formulation is used for modeling shocks.  The halos of both
galaxies in these models are represented by rigid, softened
potentials. Specifically, the halo acceleration of a particle is of
the form, 

\begin{equation}
a =
\frac{G{M_h}}{{\epsilon}^2}\ 
\frac{r/{\epsilon}}{(1 + r^2/{\epsilon}^2)^{n_h}},
\end{equation}

\noindent
where $M_h$ is a halo mass scale, $\epsilon$ is a core radius, and the
index $n_h$ specifies the compactness of the halo. When $n_h = 3/2$
the potential is like a Plummer potential, and when $n_h = 1$ the
rotation curve is flat at large radii.  The value of $n_h$ was varied,
and a value of about $n_h = 1$ was found (by experimentation with
several other values) to give the best model results for both
galaxies.  This implies flat rotation curves.  This is consistent with
the nearly flat rotation curves found for IC~2163 in \citet{elm95a}.

An adiabatic equation of state is used for the gas particles, with the
addition of a simple approximation to the standard cooling curve for a
diffuse interstellar gas of solar metallicity.  Individual particles
are heated when the local gas density exceeds a fixed threshold
density and the particle temperature is below another threshold, so
that star formation feedback is assumed to occur.  The density and
temperature thresholds are arbitrary, since the code cannot resolve
the high densities and low temperatures of real star-forming clouds.
They have been normalized to give realistic rates of star formation in
the isolated galaxies.

An approximate treatment of dynamical friction between the galaxy
halos was added to this code, because it appears that no reasonable
model of the system can be made without frictional
effects. Specifically, the Chandrasekhar formula was applied to the
rigid companion halo only, not to other particles or density
enhancements in either model galaxy. This will result in different
accelerations on particles at different radii from the companion
center and other errors, but since the frictional accelerations are
modest through most of the run, the effects are small. We take the
companion to be the galaxy corresponding to IC~2163. (To avoid
confusion about which is being discussed we will henceforth refer to
the model galaxies as galaxies A and B, corresponding to NGC~2207 and
IC~2163 respectively, and reserve the latter names for the real
galaxies.)

Specifically, we use equation 7-23 of \citet{bin87} for the frictional
force on the companion halo, treated as a point source, moving through
the primary halo, treated as an extended source.  This formula is
based on several additional simplifying approximations. First, it
assumes that $v_c/{\sigma} = \sqrt{2}$, where $v_c$ is the circular
speed relative to the fixed primary center, and $\sigma$ is the
velocity dispersion of particles in the primary halo. Both speeds are
nearly constant at large radius in an isothermal halo. In using this
formula we assume that the halo is approximately an isothermal sphere,
and that $\ v \simeq v_c,$ where $v$ is the speed of the companion
halo (relative to the primary halo). This approximation is used only
in the friction formula, though the companion's orbital velocity over
most of the orbital segment modeled here is quite close to circular,
so it is not a bad approximation in this particular system.  Thirdly,
we assume that the Coulomb logarithm term $ln{\Lambda} = 10$. The
effect of this approximate frictional force is discussed further in
Section 4.2.

There are several advantages in using this code, even though the halo
is assumed to be rigid. The first is that it requires a small amount
of computer time and memory to run, so improvements to preliminary
models can be tested efficiently. A second is that the galaxy discs
remain dynamically cold. This is in contrast to N-body codes using
modest particle numbers, where the stellar, and to some degree the
gas, discs tend to heat artificially. This is an important point in
the present case, where we need a good deal of spatial resolution in
the discs, and where we are looking at a sufficiently small part of
the system's evolution that neglect of most self-gravitational
processes (except dynamical friction) is justified. Thus, in this
paper we will focus on the results of the S97 code.

\subsection{The HYDRA Code}

Fully self-gravitating simulations were produced with the serial SPH
code Hydra 3.0 (henceforth simply Hydra), which has been made publicly
available by H.\ Couchman, P.\ Thomas, and F.\ Pearce.  In Hydra,
gravity is calculated with an adaptive particle-particle,
particle-mesh ($AP^{3}M$) algorithm (for details see \citealt{cou95},
and \citealt{pea97}).

The Hydra simulations were all run using an adiabatic equation of
state, and with optically thin radiative cooling calculated via the
tables of \citet{sut93}. These were supplied with the Hydra code. No
feedback heating was included in the Hydra simulations. Given the
limited number of particles that can be used, we cannot represent a
very extended halo with the Hydra code (unlike those in the S97
code). In any case, the Hydra models have been used primarily as a
check that the dynamical friction formulation in the S97 code gives
reasonable results for the relative orbit. This was found to be true,
in the sense that a very similar orbit was produced by the Hydra
model, with only modest changes of the initial conditions (see
Sec. 3.4).

\subsection{Scalings and Boundary Conditions}

The codes described in the previous section use dimensionless
variables.  For reference, in this section we describe conversion to
physical units.

\begin{table*}
 \centering
 \begin{minipage}{100mm}
  \caption{Parameters of the Best Model}
  \begin{tabular}{@{}cll@{}}
   & Primary Galaxy & Companion \\
   & (Galaxy A) & (Galaxy B)\\[10pt]
\underbar{Initial Galaxy Parameters}
\footnote{Physical units used in this table are derived from the
scalings: 1 code length unit = 2.0 kpc, 1 time unit = $4.0 \times
10^8$ yr, and 1 mass unit = $1.3 \times 10^{10}\ M_{\odot}$, see
text}& &\\ 

Masses ($M_\odot$):&&\\
Halo\footnote{The given halo mass is that contained within a
radius equal to that of the initial primary disc.  Gas and star disc
masses are negligible in this model, see text.}
& $1.5 \times 10^{11}$ & $1.1 \times 10^{11}$ \\
\\
Radii (kpc):&&\\
Gas Disc \footnote{There is no halo cut-off radius in the
model. Particle numbers in the best model were: A gas disc - 37200, A star
disc - 5640, B gas disc - 18090, and B star disc - 2490.}
&19.6&6.8\\
Stellar Disc&9.0&4.8\\
\\
Disc Peak Rotation (km s$^{-1}$)& 160 & 140\\
Disc Orientations\footnote{Both galaxy discs are initialized in
the x-y plane. The discs are then rotated as described. More
precisely, a statement like ``$70^{\circ}$ about z-axis'' means
$70^{\circ}$ counter-clockwise as viewed from a location at large
positive z (above the x-y plane), and analogously in all cases.}&
$25^{\circ}$ about x-axis & $55^{\circ}$ about y-axis\\
&& then\\
&$70^{\circ}$ about z-axis&$-40^{\circ}$ about z-axis\\
\\
\underbar{Orbital Parameters}\\

Initial Center Positions\footnote{Coordinates defined in a
conventional, right-handed frame, with the origin fixed at the center
of the primary. The initial separation is 23.4 kpc}& At origin &
(23.0, 0.0, 4.0)\\

((x,y,z) in kpc)&&\\ Initial Center Velocities\footnote{Relative
velocity between the galaxies is $210\ $km s$^{-1}$. Positive z
velocities are towards the observer. All disc particles were given an
initial velocity dispersion of $10\ $km s$^{-1}$} & (0.0, 0.0, 0.0) &
(-25., 210., -25.)\\ ((vx,vy,vz) in km s$^{-1}$)&&\\ \\

Center Positions when z=0 & At origin& (11.0, 19.2, 0.0)\\ Center
Velocities at z=0\footnote{Net relative velocity is $176\ $km
s$^{-1}$. The effects of dynamical friction have reduced this to less
than the initial relative velocity. The z=0 plane crossing of Galaxy B
occurs at about the same time as closest approach early in the
simulation. This is at a time of about 100 Myr after the start of the
model and about 280 Myr before the present.} & (0.0, 0.0, 0.0)&
(-170., 21. -40.)\\ \\ Gas Major Axis PA at & $\simeq 145^{\circ}$ &
$\simeq 100^{\circ}$\\ Present Time\\

\end{tabular}
\end{minipage}
\end{table*}

To scale the S97 code, we choose a velocity unit of 5.0 km s$^{-1}$.
In the code all lengths are scaled to the core radius ($\epsilon$ in
Eq. 1) in the primary halo potential, which we set to 2.0 kpc to scale
the results presented below. (While the halo core radius of the
primary is 1.0 code unit, that of the companion is 2.0 units.) The
time unit is then about $4.0 \times 10^8$ yr.  The mass contained
within a sphere of unit radius around the primary can also be
specified, and was chosen to equal $1.3 \times 10^{10} M_\odot$ here.
This gives a peak rotation rate of about $170\ $km s$^{-1}$ in the
primary disc, which is of the same order as in the HI observations for
both galaxies in \citet{elm95a}. However, the comparison is difficult
since IC~2163 is strongly affected by the tidal perturbation, and
there is evidence that the NGC~2207 disc is warped. The peak velocity
is reached at about 3.0 kpc in both models and at comparable radii in
the observations.

The S97 code dynamically extends its computational grid out to a size
large enough to include the most distant particles, or out to a fixed
maximum distance from the center of the primary.  In the present runs
this radius was set to 30 units or 60 kpc.  Few particles reached this
radius over the course of the runs; those that did were reflected
back. The timestep was also dynamically adjusted to satisfy the usual
stability requirements, but had a mean value of about $3 \times
10^{-4}$ units.

Hydra runs are made on a unit cube, where x, y, and z coordinates run
from 0.0 to 1.0 in code units. The adopted scalings for the Hydra
model are: one computational length unit equals 100 kpc, one time unit
equals $10^9$ yr, (which implies that the velocity unit is $97.7\ $km
s$^{-1}$), and one mass unit equals $10^{10} M_\odot$.

With these scalings, the results of both models suggest an interaction
time of roughly 200-400 Myr, which is measured from the time closest
approach between the two discs to the present. The large range of this
timescale reflects a liberal estimate of the range in the size of the
initial discs and the orbital separation.  Shortly after the start of
the model, this first interaction takes place on the west side of the
galaxy A disc when B moves from the near side to the far side of the A
disc relative to us (see Fig. 6). A similar timescale was deduced from
the models of \citet{elm95b}, and in fact, is likely in any model that
matches the observed rotation speeds, and has IC~2163 orbiting through
about $180^{\circ}$ (west to north to east) near the edge of the
NGC~2207 disc.

\subsection{Initial Conditions and Model Differences}

We choose the x-y plane to be equivalent to the plane of the sky in
all models. In both codes, star and gas discs were initialized by
putting a fixed number of particles in successive circular annuli,
yielding a 1/r surface density profile. The exception to this
generalization is that most models initially had a small gas hole at
their center, and then a region of constant gas/star surface density.

The initial conditions used in the S97 model are summarized in Table
3. The initial galaxy mass ratio was set to about 3/4, where NCG~2207
is represented by the larger (primary) galaxy.  The observed near
infrared luminosity ratio of about 0.6 is similar to the model
here. The peak rotation velocities are roughly comparable within the
uncertainties due to the collisional disturbances, so the
observational mass ratio constraints are not tight.

The initial midplane of the A disc is tilted by $25^{\circ}$ around
the x-axis, and then by $70^{\circ}$ around the z axis (see Table 3),
so that the near side relative to us is in the east-northeast. The B
disc is initialized with the particles first distributed smoothly in
the x-y plane, and then the disc is tilted as described in Table 3,
putting the near side relative to us in the northwest.  These angles
are constrained by the HI kinematics. The kinematic constraints are
examined in detail in Sec. 4.3.

In the Hydra simulations, the two model galaxies are identical (for
simplicity, and to get a maximal estimate of the dynamical friction)
and in both, the collisionless halo, stellar disc, and gas disc
components were added one at a time, and allowed to relax after each
addition. The halos are approximately isothermal, while the disc
components have a nearly constant vertical velocity dispersion with
radius. The discs have equal masses of stars and gas. The initial
position and velocity component values of the galaxies were assigned
such that the center of mass would remain fixed at the grid center (in
contrast to the S97 models where galaxy A is fixed), see Table 3. The
galaxies were initialized in the x-y plane after which their
orientations were adjusted exactly as in the S97 model. Initial
velocities did need to be decreased in the best Hydra models, relative
to those in the corresponding S97 model. This is to offset the effect
of the smaller halos in the Hydra models, which have less overlap, and
reduced dynamical friction relative to the S97 models.

The initial galaxies in both models were run in isolation to verify
the stability of the initial conditions. They were indeed found to be
stable, except that the gas density in the companion in the S97 model
is about equal to the feedback threshold density, so it experienced a
brief wave of star formation before settling into quiescence. 

\begin{figure*}
% \vspace{20.cm}
%\includegraphics{f4.eps}
 \caption{ Six snapshots from the fiducial model of gas particles from
galaxy B projected onto the x-y plane (i.e., plane of the sky). Galaxy
A is shown by a few representative contours of gas particle
density. In Galaxy B every sixth gas particle is plotted. In the
upper left panel, the companion trajectory through the whole
simulation is indicated by a dashed line, with a the present position
marked by a cross. Times given in Myr from the start of the
simulation; one code unit equals 400 Myr in the adopted scaling. The
axes are marked in code length units. One code length unit is 2 kpc
or about 12 arcsec in the plane of the sky in the adopted scaling.} 
\label{fig4}
\end{figure*}

\begin{figure*}
% \vspace{20.cm}
%\includegraphics{f5.eps}
\caption{Six snapshots from the fiducial model of the x-y plane, like
the previous figure, but with gas particles from galaxy A, and with
galaxy B shown by a few representative contours of gas particle
density. In Galaxy A every sixth gas particle is plotted. More
contours are also used in the later panels to outline the tail.}
\label{fig5}
\end{figure*}

\section{Results I: The Evolution of Large Scale Structure}

  In this section we will describe model results on how the large-
scale structural features of IC~2163/NGC~2207 system were formed. For
present purposes, we define large-scale structures as those that are
comparable in size to the discs of the galaxies, for example, the
ocular caustic and the tidal tail of IC~2163.  We will begin with a
general overview of the collision evolution common to the most
successful models.  In the second subsection we examine the
development of some specific structures and what constraints the
comparison of models and observation put on the galaxy structural and
collision parameters.

\subsection{General Evolution through the Collision}
\subsubsection{Evolution of the Fiducial Model}

In this subsection we describe the evolution of the most successful
model (henceforth the fiducial model) of a grid of more than 60 (S97
code) models. We also compare to another model, referred to as
Sm1. Most of the models in this grid differ only slightly from the
best ones in some orbital or structural parameters. That is, except
for a dozen or so early models this is a fairly fine grid in parameter
space located near the best models. Many of the models are essentially
the same, except for slight differences in the initial relative
position and velocity of the companion. These parameters were
fine-tuned to match the observed companion disc and tail shape and
orientation, and other bulk characteristics, such as the degree of
overlap of the two gas discs in the early encounter. Galaxy mass
ratios, disc sizes, disc orientations, and halo potential structures
were also varied.

Figures 4-7 each provide 6 snapshots each of the evolution of the two
discs in the fiducial model. Specifically, Figures 4, 5 and 6 show
only the gas particles in the discs, projected onto the x-y and x-z
planes.  Figure 7 shows the corresponding snapshots of the x-y
projection of the stellar disc. Figure 8 shows the x-y projection of
the gas disc in Model Sm1 for comparison. Model Sm1 was initialized
with smooth, azimuthally symmetric initial gas/star discs. In the
fiducial model a two-armed density wave was imposed on both the star
and gas discs in the initial conditions of the primary galaxy. More
precisely, gas/star particle positions and velocities were perturbed
from their initial circular orbit values by putting each particle on
an epicycle with phases that varied from 0 to $2\pi$ in
azimuth. Epicyclic velocity amplitudes were set to a constant times
the circular velocities, and times an $(r/{\epsilon})^{-1/3}$ fall-off
with radius. An arbitrary phase factor, varying with the inverse
radius, was also added to give the initial spiral form.  Differences
in the initial positions and velocities of the companions in the two
models with and without a spiral wave were small.

\begin{figure*}
% \vspace{20.cm}
%\includegraphics{f6.eps}
\caption{ Six snapshots from the fiducial model of the x-z plane, like the
previous figures, but with gas particles from both model galaxies.  In
Galaxy A every twelfth gas particle is plotted, and in Galaxy B every
sixth.  In the upper left panel, the companion trajectory through the
whole simulation is indicated by a dashed line, with the present
position marked by a cross.}
\label{fig6}
\end{figure*}

The fiducial simulation begins with the center of galaxy B located at
a distance of about 23 kpc away from A, and with a relative velocity
of 210 km s$^{-1}$. That is, they are on a bound relative orbit at
this point (see Sec. 4.4). The coordinate system used for the
simulations moves with the halo center of galaxy A, so that the galaxy
appears fixed at the origin throughout. The initial relative velocity
of B has a large positive y component, and relatively small, negative
x and z components. The subsequent orbit is shown in the first panel
of Figures 4 and 6.

As noted above the x-y plane is defined as the plane of the sky. The A
and B discs are initialized with the particles first distributed in
the x-y plane, and then each disc is tilted as described in Table 3.
The tilt angles are constrained by the HI kinematics, and these
constraints apply directly because the models show {\it a posteriori}
that the fundamental planes of the discs are not greatly perturbed
over the duration of the encounter. We will see below that these
choices are quite satisfactory, but detailed kinematic comparisons
suggest some improvements (See Sec. 4.3).

The first panels in Figures 4, 5, and 6 show the system at a time of
about 61 Myr from the start of the run in the representative scaling,
when the outer edge of galaxy B is just about to impact that of A from
the near side (relative to us, i.e., at positive z). Closest approach
is at a time of about $t = 0.25$ units or 100 Myr; see the second
frame in the figures.  This disc collision on the north and western
side of A continues through the time of the third frame in these
figures.  During this time the two discs scrape against each other
across a strong shock front.  This scraping radially compresses the
effected portions of the two discs.  Another effect is that some gas
from B is transferred to the A disc (see Fig. 4).

Note that we are not completely certain that this disc collision does
in fact occur.  A slight change in the initial conditions could lead
to either no impact between the gas discs or greater overlap and a
much more widespread and destructive impact. We have run models of
both situations. As a result of those runs we believe that encounters
with much greater overlap lead to NGC~2207 disc morphologies at the
present time in the south and west that are very different from those
observed. The possibility of a complete miss is harder to rule out,
but several observations support the idea of an early disc collision
like that shown here.  The primary one is that the NGC~2207 disc does
appear disturbed in the south and west.  For example, as we will see
in the next section, most of the small scale peculiar emission regions
are located in this area. This issue is discussed further in Sec. 4.3.

The last four panels of Figures 4 and 5 show the companion swinging
roughly another $90^{\circ}$ in a counterclockwise direction around
galaxy A as time advances from 180 Myr to 380 Myr. That is, it takes
about 380 Myr for the companion to move from the west to east side of
the A disc. The specific ocular and tailed morphology of IC~2163 are
relatively short-lived (see discussion in Sec. 6 for specifics). If
the disc collision did occur, then it is likely that the x-y part of
this model trajectory is reasonably correct. A very different
trajectory would not get the companion to its present position at the
time when these morphologies are present.

\begin{figure*}
% \vspace{20.cm}
%\includegraphics{f7.eps}
\caption{ Six snapshots from the fiducial model of the distribution of
disc stars in the x-y plane. All star particles from both model
galaxies are shown. Note we use a magnified scale relative to previous
figures in order to show the development of the ocular structure in
galaxy B and the density wave structure in galaxy
A.}
\label{fig7}
\end{figure*}

\begin{figure*}
% \vspace{20.cm}
%\includegraphics{f8.eps}
\caption{Four snapshots of the distribution of gas particles in the
x-y plane for model Sm1 without an initially imposed spiral,
for comparison with Figs. 4 and 5.}
\label{fig8}
\end{figure*}

For example, if the galaxy B orbit had substantially less momentum or
angular momentum, then it would travel under (below), but closer to
the center of the A disc. There would have been more interaction
between discs, and more dynamical friction between the haloes. As a
result the B disc would be unlikely to emerge on the east side; it
would complete the merger promptly. On the other hand, in many of our
early runs the B orbit had more angular momentum, and B moved much
farther from the A disc after the disc collision. Specifically, the
separation between discs was much greater than observed when a
distinct ocular/tail morphology was present.

It is possible that the z-motion was greater, and that the galaxies
have a greater separation along the line of sight. This would require
that the two galaxies be near apoapse at present, to reduce the line
of sight velocity to the observed low level. However, if that were the
case, the total velocity (and its x, y components) could not be much
reduced, or the collision would be more direct, with more damage to
both galaxies, and there would be a ring galaxy remnant in place of
NGC~2207.  We believe that both the disc collision and the relative
orbit of Figures 4-6 are quite well constrained, and will discuss these
constraints more quantitatively in Section 4.2 below.

The fiducial model was run another 200 Myr beyond the present. In that
time the companion moves down and inward a bit, and then spirals
in to begin the final merger with the primary disc.

Figure 7 shows the distribution of disc stars from both galaxies in
the x-y plane, but on a magnified scale relative to that of Figures 4
and 5. Comparison shows some significant differences in the evolution
of the gas and star discs. First of all, the stellar oculars in the
panels of Fig. 7 are much sharper and more prominent than the gaseous
oculars in Fig. 4. In general, the star discs are less disrupted by
the encounter than the gas discs, which explains many of the
differences in the wave structure in the gas and stars (though the
initial stellar volocity dispersion also determines stellar wave
sharpness). A related difference is that there is minimal mass
transfer between the stellar discs, while there is a good deal of gas
transfer. The gas discs also experience strong shock waves, and
transfer is helped by the fact that the initial gas discs are larger
than the star discs. Indeed the stellar disc of the primary looks
hardly disturbed at the last time shown in Figure 7.

This concludes our overview of the fiducial model. Next we consider
the origin of the spiral structure in NGC~2207.

\subsubsection{Initial Spirals Versus Induced Waves}

Figure 8 displays the evolution of the system in model Sm1, which has
very similar initial parameters as the fiducial model, but without the
initial spiral.  The figure shows that the gas disc of galaxy A
appears only moderately disturbed by the interaction with B, and
specifically, there is no evidence that the beautiful two-armed spiral
of NGC~2207 can be produced in such a collision. The many other models
we computed affirm this conclusion. The basic reason for this is
clear; the companion orbits in a retrograde sense relative to the A
disc rotation. It has been known since \citet{too72} that prograde
collisions produce strong M51 type two-armed spirals, but retrograde
generally do not. Thus, we believe that the spirals predated the
current interaction in this system.

This hypothesis raises the questions of whether pre-existing spirals
would survive the interaction, and if so, how would they be affected?
The last frames of Figures 5 and 7 suggest that they are not greatly
affected, which is consistent with the impression from Figure 8 that
the A disc as a whole is not highly disturbed. (Note that the imposed
kinematic spiral persists for several rotation periods in isolation,
so it is not amplified by the collision either. We should caution,
however, these conclusions may be modified by models with full disc
self-gravity (rather than the local self-gravity of the S97 code).)

Similarly, in \citealt{elm00a} we suggested that some of the filamentary
structure in the IC~2163 tail might have originated in flocculent
spiral arms before the collision. The present result seems to support
the idea that flocculent arms would persist in some form.

\subsection{Timing and Orbital Constraints From Large Scale Structure}

In this section we describe how the presence of several short-lived
tidal structures that are sensitive to the details of the
time-dependent perturbation, and thus the relative orbit and other
collision parameters, allow us to constrain these latter parameters
unusually well. The strongest constraints are provided by the
structure of the IC~2163 tail, the presence and orientation of the
IC~2163 ocular feature, and the structure of the extended HI disc in
NGC~2207. We will consider each of these in turn, and then briefly
consider the accuracy of the dynamical friction approximation used in
the S97 models.

In contrast to many galaxies with tidal tails in the \citet{arp66} and
\citet{arp87} atlases, the IC~2163 stellar tail is not very much
longer than the mean disc diameter of that galaxy. This stellar tail
also has an unusually large width to length ratio. Both gas and
stellar tails get longer and narrower as they develop (see Figs. 4 and
8). Thus, the tail structure argues for a relatively young age.

In fact the gas tail in the last panel of Fig. 4 looks a bit too long
and narrow to match the observations of Figs. 1 and 3. However, Figure
8 in \citet{elm95a} shows that the gas tail is longer than shown in
Fig. 3. Moreover, the last panel of Fig. 7 (and other larger scale
views) shows a stellar tail that is generally like that observed,
though it looks more like the observation at a slightly earlier time.

The difference between gas and star tails also shows that tail
structure is sensitive to the initial size of the disc, with larger
discs developing tails earlier. On the other hand, the size of the IC
2163 disc relative to that of NGC~2207 is constrained by
observation. Taking these factors into account we estimate that the
``tail age'' of this model is in the range 220-260 Myr after closest
approach, or 320-360 Myr from the start of the model. Note, however,
that time estimates in this section could be increased slightly and
decreased by as much as 100 Myr if the model galaxy sizes and orbit
sizes were varied over the maximal range allowed by observation. 

Next we consider the ocular structure, which is best seen in the star
particles shown in Fig. 7. Specifically, if we use the rather liberal
definition of an ocular as a pointed oval, we get a time range of
about 60-260 Myr since closest approach (the early disc collision).
However, at the beginning of this time range the ocular is not in the
correct position. At the end of this range, it is developing a rather
different appearance as the primary stellar ocular disappears and a
second ring-like wave appears. In the middle of this range, the ocular
looks quite like what is observed, with sharp (bright) rims on the top
and bottom. Also at these times the angle between the upper tail and
the upper rim is like that observed.

The IC2163 ocular and tail are two parts of a single tidal distortion;
to get a different timing estimate we must look at the NGC~2207
disc. Since we have argued that the prominent spirals in this disc are
not a result of the current interaction, neither these spirals, nor
any other large-scale optical structure, provides such
information. However, comparison of the elongated oval HI distribution
in Figure 2 to the model gas distribution of Figure 5 suggests that
the shape and position angle of the major axis of the gas distribution
can give an age estimate.

Specifically, the observed angle between the HI major axis of NGC~2207
and the line connecting the galaxy centres is about $45^{\circ}$. This
angle is approximately matched in the models at times in the range
240-280 Myr (since closest approach, or 340-380 in Fig. 4 or 5), like
the previous estimates. It appears that the gas in the models does not
extend as far to the south of the center of galaxy A, as in the
observations. However, we note that the existence of this oval
distortion, and the time taken to form it, strongly suggests a
prolonged encounter, like that in the model. Finally, the last panel
of Figure 4 shows that some gas transferred from the companion is
found in the southern extension.

Because of the approximate treatment of dynamical friction in the S97
code we should emphasize again that the observational constraints do
not seem to allow a great deal of deviation. Secondly, though
important, the effects of friction up to this point in the system's
evolution are modest. The models suggest much stronger effects in the
immediate future. Thirdly, because the encounter is retrograde
relative to the rotation of NGC~2207, and because the halo of the
companion is more massive than the disc of NGC~2207, we do not expect
strong couplings between the companion and resonant disc orbits, which
could modify dynamical friction relative to the Chandrasekhar
approximation \citep{tre84}. Fourthly, we have produced similiar
orbits to the best S97 models with the fully self-consistent Hydra
code. However, because of the different halo structures in the two
models, we have not attempted to produce a detailed match with the two
codes.

\subsection{Kinematics}

\begin{figure*}
%\scalebox{0.99}{\includegraphics{f9.eps}}
\caption{Observational (top two panels) and model (lower four panels)
position-velocity diagrams. In the observational panels the emission
is summed over all relevant values of right ascension (top left) and
declination (top right) using data from \citet{elm95a}. The emission
from the two galaxies has been separated, with that from NGC~2207
shown as contours, and that from IC~2163 shown as gray-scale. In the
model gas particle position-velocity diagrams Galaxies A and B are
shown separately in the middle and bottom rows,
respectively. Coordinates like declination and right ascension are derived
from code units by using the adopted scale length and
distance. Specifically, we take 1 length unit (2 kpc) = 0.2
arcmin. (12'') of declination = 0.84 seconds of right ascension. Note
that negative numbers are used for offsets to the east (left). For
explanation of the labeled features, see text. }
\label{fig9}
\end{figure*}

\begin{figure*}
%\scalebox{0.99}{\includegraphics{f10.eps}}
\caption{Gas and star distributions from the observations (left
column) and the fiducial model (right column) at about the present
time (t = 380 Myr) are shown for reference. Specifially, the upper
left panel is an optical image (from the Digital Sky Survey), the
lower left panel is a greyscale total HI image (data from
\citet{elm95a}), the upper right pamel shows model stars (in the inner
disc of Galaxy A) and star-forming gas particles (i.e., young star
clusters often located in spiral arm segments), and the lower right
panel shows every second gas particle. All figures are shown on
comparable scales. Model coordinates with units like declination and
right ascension are derived from code units as in the previous figure,
observational coordinates are offsets from the nucleus of NGC~2207.}
\label{fig10}
\end{figure*}

Models for this system are quite well constrained by the various tidal
morphologies. However, comparison of models to the line-of-sight
kinematics provides important checks, and additional input on collision
parameter values to which kinematic structures are especially
sensitive. These latter include disc orientations and warps.

The observed HI kinematics are summarized in the two observational
position-velocity diagrams of Figure 9. Figure 10 shows gas and
stellar distributions for comparison, and for reference in the
following discussion.  In the top left panel the emission summed over
all relevant values of right ascension is displayed; in the top right
panel, the emission is summed over all relevant values of
declination. These diagrams reveal the extent to which the two
galaxies overlap in line-of-sight velocity.

We consider first the summed P-V diagrams for IC~2163. The large
velocity range in IC~2163 results from velocity streaming along the
ocular oval and along the tidal tail and tidal bridge. In the top
left panel, the two sides of the rim of the ocular oval are distinct,
and the tidal tail of IC~2163 appears as a constant velocity feature
at the highest velocity (v = 3000 km s$^{-1}$) extending farthest to
the north. In the top right panel, the tidal tail is the wide, constant
velocity feature (v = 2900-3000 km s$^{-1}$) extending farthest to
the east; its 100~ km~ s$^{-1}$ velocity width in the RA-V diagram is a
result of gas streaming outwards along the inside edge of the tail and
inwards along the outside edge of the tail (see \citealt{elm00a}). The
IC~2163 gas at v $< 2640\ $km s$^{-1}$ is mainly tidal bridge gas.
The brightest features in these P-V diagrams of IC~2163 are produced
by the massive HI clouds.

In the summed P-V diagrams of NGC~2207, the HI southern extension is
clearly seen as the nearly constant velocity (v = $2600-2680\ $km
s$^{-1}$) feature extending farthest to the south in the top left
panel of Fig. 9 and as the eastern low-velocity clump in the top right
panel. Much of the emission in the far north at a velocity of 2875 km
s$^{-1}$ is associated with HI cloud N1 (Figure 2).  The encounter
model in \citet{elm95b} suggested that a strong warp and perpendicular
motions set in at about the distance of cloud N1.

The kinematics of the fiducial model are summarized by the four summed
position-velocity diagrams in the lower four panels of Figure 9, which
show the Z component (line-of-sight) of the velocity versus X and Y
coordinates for the gas particles in each of the two galaxies. We have
attempted to facilitate the comparison of these plots to the
observational plots by using the arcsec-to-pc conversion of note b) of
Table 1 and the model scalings of Section 3.3 to transform code units
to right ascension and declination. The specific scaling values are
given in the caption.

Comparison of the observational and model figures reveals that most of
the large-scale kinematic structures are similar, with a few
exceptions. We first consider the right ascension plots. In both the
model and observation plots the companion is located to the east of
the primary, with some overlap. The primary is generally oval-shaped
in these plots. Both model and observation plots show emission peaks
on the high and low velocity sides of the oval. In top right panel of
Fig. 9 the companion emission has a faint, thin, tilted distribution,
with a strong extension to the east at large velocity ($2900-3000\ $km
s$^{-1}$) due to the tail. The strongest emission in the model plot
has the same general characteristics, with a couple of differences.

The first is that the model disc B itself has a few gas particles
extending to higher velocities ($> 2900\ $km s$^{-1}$) on the west
side (r.a. $>\ -5.0$), and also much more diffuse gas at intermediate
velocities. The intermediate velocity gas is found to be material
accreted onto the primary disc. A significant fraction of this gas is
in warm phase. We should also point out that much of this gas is
probably counted as part of the primary galaxy in the
observations. Merging it with the primary in the model declination
plot would make that plot look more like the observations. The higher
velocity gas is located, on average, slightly above the mean disc
plane (in the positive z direction). These high velocity particles
were scattered in the scraping interaction between the two discs, but
not accreted. The absence of the material in the observational plot
may indicate that this gas was accreted onto the primary in the real
system, or that this material is below the observational sensitivity.

Next we consider the declination plots. In these plots the centres of
the two discs generally overlap and the extent of each is much less
than in right ascension. For the primary, most of the model particles
and much of the observed emission are located on a nearly linearly
rising velocity curve. This feature rises slightly more steeply in the
model than in the data. At low declination in the observations, there
is a nearly constant velocity extension to the south. On the sky this
corresponds to the southern HI extension seen in Fig. 2. This feature
(marked as the southern extension ``Sp'' (for spiral) in Fig. 9) is
weak in the model. It consists entirely of particles in the spiral arm
that extends to the southeast in the last panel of Figure 5.

The middle left hand plot of Fig. 9 shows that the model primary has
two extraplanar plumes that are not seen in the observations. The
first of these is labeled `NArm' in that plot and the corresponding
right ascension plot.  This material is located in scattered parts of
the northeast quadrant of the primary, and much of it is
extraplanar. The disturbed velocities of these particles originated
from the turn-on of feedback effects in the southern spiral arm at the
beginning of the run. Since this initial transient SF and feedback is
unrealistic, they are a numerical artifact.

Particles at velocities less than about 2600 km s$^{-1}$ in the middle
left plot of Fig. 9 (marked ``Scrape'') make up a kind of diffuse
extension of the southernmost part of the primary disc. These
particles, and the diffuse particles between the ``Sp'' and ``Scrape''
loci, come from three sources. First, some are products of spiral arm
feedback, and thus, are artifacts of initial transient effects in the
model, like the NArm. Second, some are primary disc particles that
were located in the northwest at the start of the simulation, and were
perturbed as a result of a scraping interaction with particles in the
companion gas disc. The third group of particles come from the
companion, and were on the other side (east side) of the scraping
interaction. These particles were removed from the companion and
orbited under the primary disc to their present location.

In the real system, as shown in Fig. 9, there is essentially no
primary gas at velocities of less than 2570 km s$^{-1}$. The most
likely explanation for the difference is that the scraping interaction
between discs may be too strong in the model.

Next we consider the companion declination-velocity
plot. Observationally, the companion looks like a thin, nearly
vertical oval, with a faint horizontal plume attached in Fig. 9. The
model image, in the lower left plot of Fig. 9, is very similar,
though compressed (appearing skinny). The shape and orientation of
this oval are sensitive functions of the B disc orientation. The
initial B disc orientation is close to correct, but not perfect.

In the model, particles making up the horizontal plume are found in
two distinct regions. Most of the particles come from the companion
tail. The remainder are scattered through a region in the northern
part of the primary galaxy, though they originated in the
companion. They were accreted in the first encounter between the two
discs. They have never had any real association with the tail
particles.

In sum, the model can account for all the major kinematic features in
this system, except for warps in the gas disc of NGC~2207, and the
detailed structure of the southern HI extension of that disc. In the
latter case, the model does in fact have particle features
corresponding to that structure, but because of the effects of
particle resolution and spiral arm parameters they do not match the
observed kinematics especially well. The model also makes predictions
about the kinematics of accreted gas, but these will be hard to test
observationally.

\subsection{Before the Present Encounter}

Before leaving the topic of overall evolution to discuss finer
structural details, we should consider the question of what might have
happened before the beginning of the fiducial model. Recall that the
fiducial model begins with the companion located a short distance out
from the initial closest approach point to the primary, and that after
the close approach the companion continues around on a nearly circular
orbit. It is natural to infer that this orbit is the continuation of a
prolonged nearly circular orbit or one which brings the companion
slowly spiraling inward. If so, the tidal interaction would be more
prolonged than indicated by the fiducial model, and this, together
with previous close encounters, could have a strong effect on the
system not accounted for by the fiducial model. (E.g., see the models
of prolonged interaction in the M51 system by \citealt{sal00a,
sal00b})

To test this possibility we ran the fiducial model 'backwards' from
its starting point by reversing the initial velocity of the
companion's relative orbit, and reversing the sign of the dynamical
friction term. We found that, indeed, the relative orbit of the
companion is qualitatively like a precessing ellipse with modest
eccentricity. It's furthest radial excursion was only about 50\%
farther out than its closest approach radius.

Given the companion's relative position and velocity at a time about
equal to one orbital period before the beginning of the fiducial
model, we ran the model forward from that time to judge the
effects. The primary was little affected by the prolonged, but still
retrograde encounter. However, as would be expected, the effect on the
prograde companion was strong. The tidal tail (and bridge) developed
immediately, and the tail was very long by the time when the fiducial
model was started. By the present time this tail would bear no
resemblance to the much less mature observed tail. Thus, there is
strong evidence against a prolonged encounter, and the "pre-history"
of the fiducial model must be modified from the simplest time reversed
extension. 

Of the many possible changes we could make to the relative orbit in
the fiducial model to avoid the prolonged encounter, a great many can
be eliminated because they would lead to a model result that does not
match the observations nearly as well. In fact, we don't really want
to change the orbit of the fiducial model itself, but only the orbit
before the starting point of that model. To match the observations we
want the companion to move well away from the primary as we go back in
time, and to be well away for a long time before the present.

This scenario plays out very naturally if the nearly flat rotation
curve halo of the primary does not extend much farther out than the
radius of the companion at the start of the fiducial model. We set up
a model with the primary halo density declining beyond this radius,
such that the circular velocity fell off as $r^{-0.25}$ (i.e., midway
between a flat rotation curve and a Keplerian curve). When this model
is run backwards from the fiducial start conditions, the companion
flys off to large distances. This example, and the observational
constraints, suggest that the companion was initially on a more nearly
parabolic orbit of relatively low angular momentum, so that it was
able to move into the outer parts of a modest-sized primary halo and
be captured onto an orbit like that of the fiducial model.

It may seem that we have to fine-tune the structure of the primary
halo to get this solution, but it seems to be required to match the
observations. Moreover, the precise density drop in the outer halo
is not strongly constrained, and thus, not finely tuned. Finally, this
fall-off can help account for the southern gas extension, which is not
reproduced well in the fiducial model. Gas that is swung out to the
radius where the halo density declines more rapidly, will not fall
back as quickly, and may in fact travel farther out than in the
fiducial model with no halo cut-off. 

\section{Results II: Development of Fine Structure} 

In this section we look somewhat more carefully at a few of the
collisional structures of this system, and their origin. Both the high
quality of the multi-wavelength observations of this nearby system,
and the detailed reproduction of its major features by the models,
argue that it is worth pushing the comparison of them to a higher
level of detail than usual. In fact, we find that we can give a
plausible account of the development of the features considered, and
also provide predictions to be checked by future observations.

The overall structure of the IC~2163 bridge and tail are generally
what we would expect for a prolonged prograde encounter at an
intermediate stage of development. However, there are some
peculiarities (see Tables 1 and 2). The bridge, lying behind the
NGC~2207 disc, is difficult to discern in the optical beyond the
western cusp of the ocular feature. A longer bridge is seen in the HI
(see Fig. 3). At later times the models show a good deal of gas has
been or is being transferred to A (Fig. 4).  Cloud I4 in Fig. 3 may be
a result of such ongoing transfer. This gas is connected to the B disc
by a very broad, diffuse gas bridge (Figs. 4, 6), which would be hard
to identify in the HI observations aside from its contribution to the
observed high velocity dispersion in the HI gas. The fiducial model
has a substantial gas bridge (Fig. 4 or 5), which looks rather similar
to the observed HI bridge. Most of our other models do not reproduce
this feature, but the fiducial model has the largest initial gas
disc. The fiducial model stellar bridge is significantly offset from
the gas bridge and has at best a faint counterpart in the
observations. This may mean that the initial model star disc is too
large.

The observed stellar tail is rather short and wide compared to many
in, for example, the \citet{arp66} atlas. However, the models suggest
that this is mainly a result of its youth. The tail has a great deal
of internal structure, including a number of dark dust and bright
stellar filaments.

\begin{figure*}
%\scalebox{0.99}{f11.eps}
\caption{Four snapshots of the evolution of two circles (ellipses in
projection) of gas particles in Galaxy B. In these plots the origin of
the coordinate system has been moved to the potential center of Galaxy
B, which is marked by a cross. The positions of groups of 4 particles
were averaged in order to smooth distortions due to single
particle scattering. See text for details.}
\label{fig11}
\end{figure*}

\begin{figure*}
%\scalebox{0.99}{\includegraphics{f12.eps}}
\caption{Two x-y snapshots of selected gas particles (shown as crosses
in the top row) apparently transferred from Galaxy B to Galaxy A. The
top right-hand plot corresponds to about the present time ($t = 380$
Myr). The left-hand plot of the first row shows the
same particles at the beginning of the run, illustrating where the
captured particles originated. The second row shows two views, x-y and
x-z, of the trajectories of 10 captured gas particles. 10\% of all the
gas particles are shown as small dots in all panels for reference,
except for the first panel where every fourth particle is plotted.}
\label{fig12}
\end{figure*}

Figures 4 and 6-8 provide a good view of the overall development of
the model tail. However, they provide only rather general information
on the origin of tail and its fine structure. Figure 11 provides more
details, with snapshots of the evolution of two rings of gas particles
in the initial Galaxy B. By the second time shown we can already see
the effects of tidal stretching in one direction, and compression in
the orthogonal direction. There also appear to be overall compressions
and expansions of Galaxy B associated with the ocular waves and the
angular momentum perturbations.

The gas particles in these rings are also affected by hydrodynamic
shocks. For example, shocks in the disc-disc interactions probably
give rise to some of the jagged structure in Fig. 11. In fact, some of
the ``bridge'' particles in the lowermost part of the last three
snapshots have likely been captured by galaxy A.

While all the particles in each ring were initially on near circular
orbits with nearly the same period, the tidal forces change the orbits
to ellipses with phase and period gradients. As a result, ring
segments cross other segments in both the same and other rings, as is
evident in the final two snapshots of Fig. 11. Many of these crossings
occur in parts of the ring that have been pulled out into the tail. To
the extent that gas in the tail remains coplanar, shocks will result
from attempted orbit crossings (in contrast to the stars). The sizes
of ring self-crossings in the last three snapshots of Fig. 11 are
similar to those of the dust segments in the HST imagery of the
IC~2163 tail (see Fig. 1 and \citealt{elm00a}, \citealt{elm00b},
\citealt{elm01}).  Thus, these segments may be explained by the
relatively small scale shocks produced by such crossings (also see the
model of \citealt{elm00a} and the discussion in \citealt{sal93}
related to their Fig. 8).

Assuming particles at different initial radii have different
metallicities, this process would mix stars and gas clouds of
different metallicities, especially in the inner tail. The outer parts
of the tail would consist mostly of metal-poor gas from the outer
parts of the initial disc.

Figure 12 provides a more detailed view of the mass transfer of gas
from Galaxy B to A. The upper right panel of this figure shows a
sample of gas particles (plus signs) at about the present time, which
originated in B and which have been captured by A. Specifically, the
plus signs mark 10\% of the B particles with $x > -3.0$. This is not a
rigorous capture criterion, but since it generally requires the
particles to be closer to the potential center of the more massive
Galaxy A than to B, it is generally an effective one.

The upper left panel shows the same particles at the onset of the
simulation. The plus signs are all clumped in the upper half of the B
disc, which is the region that interacts most strongly with the disc
of A (see Fig. 6).

The lower two panels of Fig. 12 show x-y and x-z views of the
trajectories of 10 representative captured particles, from the start of
the simulation up to the present time. These panels make clear that
the disc-disc interaction scatters affected gas particles quite
widely. Some generally continue to orbit between the two galaxies,
though now apparently perturbed out of the plane of the B disc. Others
are stopped and have the sign of their angular momentum reversed, so
that they orbit in the same sense as particles in the A disc (though
again in different planes). 

Because of their different orbital planes, these particles pierce the
gas disc of A, probably inducing strong cloud collisions, and local
heating. At the present time, their distribution is weighted towards
the southeast quadrant of the A disc. Apparently, there has only been
sufficient time for the accreted material to orbit to that part of the
potential.

\citet{elm00a} noted about 20 peculiar emission features in the HST
observations of this system (see Fig. 1). Of those that appear to
originate in NGC~2207, most are located in the southern half of that
disc. It seems quite possible that a number of these were produced by
gas clouds from IC~2163 with trajectories like those in the lower
panels of Fig. 12. That is, trajectories that are perturbed by
interactions with gas clouds in the northern part of the NGC~2207
disc, pierce that disc, and subsequently return to collide with it
again in the southern half. (We caution, however, that this depends
sensitively on the companion's trajectory.) The wide dispersion of the
model trajectories is certainly in accord with the dispersed locations
of the emission regions. Several of the emission regions have linear
or arc-like forms that could be produced by cloud collision bow
shocks.

The far western emission region {\it{i}} is also coincident with a
strong radio continuum source, and two candidate super-star
clusters. This region may have been directly excited, or driven into
vigorous star formation, by the disc-disc interaction. In the fiducial
model, the disc-disc interaction occurs most strongly in the
northwest, and by the present time, the affected material has rotated
to the southeast. If the disc-disc interaction either occurred at a
later time along the orbit, or continued longer, so that it was still
underway when the companion passed from the north to the northeast of
the primary disc, then that part of the primary disc would now be
located on the western side of the NGC~2207 disc, near region
{\it{i}}.

Alternatively, the activity of region {\it{i}} could be stimulated by a
milder interaction than the disc-disc encounter that captured so much
material from the companion. The models show that after this strong
interaction, the near side of the companion disc continues to press
the passing spiral arm of NGC~2207 inward.

Either this prolonged scraping interaction, or continued mass transfer
from the bridge, may be responsible for producing the ridge of radio
continuum emission on the northeast side of the NGC~2207, described in
\citet{elm95a}. Additional observations and analysis (to be published
elsewhere) have shown that the continuum ridge is in fact coincident
with the middle spiral arm in the northeast of NGC~2207. This result
would seem to favor interaction between that arm and IC~2163 as the
cause of the emission. Changes in the spectral index along the ridge,
indicative of an aging, cooling population of radiating electrons,
would provide some confirmation for this hypothesis, as opposed to the
possibility of a random collection of local point sources.

\begin{figure*}
%\scalebox{0.99}{\includegraphics {f13.eps}}
\caption{x-y snapshots of star-forming gas particles (shown as plus
signs), with 10\% of all the gas particles shown as small dots for
reference (except in the lower right panel where 20\% of the particles
are shown). The upper left-hand panel shows an early time near the
beginning of the run. The lower left-hand plot shows SF in the
developing first ocular wave of Galaxy B at a later time. By the time
shown in the upper right panel, the ocular wave has nearly propagated
out of B, and a new wave is forming and triggering SF. The lower right
panel shows a time near the present, with a magnified scale, focussing
on the companion; for more details see discussion in text.}
\label{fig13}
\end{figure*}

\section{Spatial/Temporal Patterns of Star Formation and
ISM Phases}

The S97 code contains a simple SF and feedback algorithm as noted in
Section 3.1 and described in more detail in \citet{str97}. SF is not
particularly strong in this system, and our treatment of it in the
simulation code is very approximate, so we should expect that the
information provided by the models on this topic will be
limited. Nonetheless, the large scale spatial pattern of SF in these
two galaxies is interesting, and it is of interest to see how well the
models can match it, and what they have to say about its origin and
history.

Figure 13 provides four snapshots of the pattern of SF in the fiducial
model at various times, with the last at close to the present time.
The upper left panel of Fig. 13 shows a time near closest approach,
when the two gas discs have begun to scrape against each other. The
resulting compression induces a small and brief SF enhancement in the
adjacent part of the Galaxy B disc in some of our models. In the
fiducial model the scraping involves a spiral arm segment which
experiences enhanced SF as a result of the compression. Other parts of
the spiral are not particularly active, and Fig. 14 shows little
excess SF in the A disc at this time. A core SF burst is beginning in
the B disc as the first ocular wave begins to develop. The spiral arms
are created in the initial conditions with relatively high gas
densities, which generate strong star formation at the beginning of
the simulation. This transient effect has largely damped by the time
shown in the first panel. It is responsible for the burst at early
times shown by the solid curve in Fig. 14.

The SF in the Galaxy A disc has damped to a more typical level by the
time shown in the second panel. By this time the ocular wave has begun
to develop in the (displaced) center of Galaxy B, and the resultant
compression drives stronger SF. Up to this time, the Galaxy B disc had
generated very little SF (except in an early transient ring wave),
since it was initialized with gas densities below the
threshold. Scattered SF particles are also seen in the nascent tidal
tail at this time. There are too few particles involved to attach much
significance to this. However, the general pattern is typical of all
timesteps after the onset of tail formation, and reasonable, since we
expect local gas compressions in the tail. The absence of SF in bridge
gas is also interesting. A large fraction of these gas particles have
relatively high temperatures (due to the disc-disc interaction) at
this time.

The third panel of Fig. 13 shows a time when the ocular wave has
nearly propagated through the B disc. Compare this to Fig. 7, which
provides better views of wave evolution in the companion. At about
this time, a second oval wave appears in the center of the companion
and begins to propagate outward (see Fig. 7). As in collisional ring
galaxies, this second wave is primarily a product of continuing
coherent radial motions resulting from the earlier perturbation, not
any new disturbance. This coherence is somewhat less in the gas than
in the stellar disc.

The fourth panel shows a time near the present, where the second
ring-like wave has propagated well out in the remaining companion
disc. The view in this panel is magnified to show the SF in the
companion, which is relatively strong and concentrated in the
wave. The SF is strong in the upper rim of the oval, as
observed. However, examination of many other previous timesteps shows
that the SF moves around the oval waves as they propagate, so the
location of SF in the last panel is a short-lived situation, and the
coincidence with observation is somewhat fortuitous.

The first three panels of Fig. 13 show that SF in Galaxy A is
concentrated in the spiral arms. This is in qualitative agreement with
the observational results of \citet{elm01}, in which the most luminous
young star clusters in NGC~2207, and candidate super star clusters,
were found in various parts of the spiral arms.

Figure 14 shows the total amount of SF in each of the model discs as a
function of time in the fiducial and for the companion in other
models. The SF history in the B disc is qualitatively similar in the
models shown. The first peak is associated with the first ocular wave,
and exceeds the SF at all other times in either disc (excluding the
first, artificial peak in disc A). The time interval between the burst
and the present is about 260 Myr (see Figs. 4 and 5). This suggests
that a post-starburst population might be present in IC~2163. If
observed, this population would provide some circumstantial evidence
for the early compression associated with the formation of the ocular
waves. The rapid rise in SF at the latest times (i.e., in the future)
is due to the onset of merging.

\section{Discussion and Summary}

Motivated by the proximity of the NGC~2207/IC~2163 system, and the
high quality and resolution of the multi-waveband observations, we
have attempted to model its morphological and kinematic structure in
considerable detail. The collisional models described above have
successfully accounted for the major, large scale structures in this
system. In this section we review Tables 1 and 2, in order to see
which of the 34 features listed there are accounted for by the
models. We also want to consider which features might be so sensitive
to specific model parameters that, while they may, accidently, appear
in a particular model, they are unlikely to be produced in a generic
model. In striving for accurate models, can we begin to perceive the
limits of such modeling?

We begin with items 1 and 4 of Table 1. The distance and mean redshift
are not relevant to the models, though the difference between the LOS
velocities constrains the relative orbit. Item 2 gives luminosity
ratios. Table 3 shows that the halo mass ratio (0.7) is quite similar
to the observed NIR luminosity ratio (0.6). This similarity may be
fortuitous unless the halo mass ratio of the two galaxies is the same
as their old star mass ratio. If the mass ratio were much lower the
perturbation would not be strong enough to perturb the Galaxy A gas
(HI) disc, as observed. If it were much higher, the Galaxy A disc
would be more disturbed.

\begin{figure*}
%\scalebox{0.5}{\includegraphics{f14.eps}}
\caption{Total number of star-forming particles versus time within the
model galaxy discs. The solid curve shows the SF in Galaxy A of the
fiducial model. The dashed curve shows the SF in Galaxy B of that
model. The dotted and dash-dot curves show the SF in companion Galaxy
B in two different, but quite similar models. Comparison of these
Galaxy B curves allows a determination of which features are generic;
see text.}
\label{fig14}
\end{figure*}

The young star luminosity ratio is quite different (0.3), though more
young stars may be obscurred in IC~2163 than in NGC~2207, since in the
latter case the SF regions are generally well outside the
core. Fig. 14 shows that the ratio of SF particles in the two galaxies
varies greatly with time. At the present time it is roughly
unity. Simple feedback formalisms can only be expected to give
qualitative information about the SFR.

Items 3, 6, and 10 concern radii and separations. First of all, we
note that the radius of IC~2163 in item 3, from the RC3 catalog, seems
to include the tidal tail. The initial radii of the old star discs in
the model galaxies are not as well constrained as gas disc radii are
by the need to match tidal morphologies. They can be changed over a
modest range with little effect on the collisional or spiral arm
morphology. A third caveat is that the ocular diameter of item 10 is
time-dependent, as the ocular waves evolve. Despite these caveats the
sizes and separations of the fiducial model seem about right (Figs. 4,
5, 7 and 10).

Item 5 describes the observational (HI) estimate of the orientation of
the two discs. In the case of IC~2163 the models agree with this
estimate. In the case of NGC~2207, the nearest side in the models is
in the east, while the observations suggest it is in the NE to
N. However, the models have little of the disc warp indicated in the
observations (item 24). Moreover, the observed HI velocity field could
not be fit by a static, tilted ring model; extra z-motions had to be
included. Cold gas accreting onto NGC~2207 could contribute to these
extra z-motions, as well as tidal forces.

Item 7 notes the typical HI mass to SFR ratio. The SF terms in the
model yield qualitative agreement with the observations on the
distribution of SF. Item 11 notes enhanced emission in the ocular rim,
and the SF model also produces this (Fig. 13).

Item 8 notes widespread, high velocity dispersion HI gas. We have not
examined the model gas velocity dispersion in detail. However, Fig. 9
shows the presence of a good deal of warm gas with a large velocity
dispersion (e.g., accreting gas in the eastern half of the lower right
panel of Fig. 9). This gas also has a large range of temperatures,
and is probably only partially ionized, so may correspond to some of
the observed high dispersion gas.

Item 12 describes the major axis orientations of IC~2163. In the
models the Galaxy B orientation is time-dependent, but at the present
time (t = 270 Myr), Fig. 7 shows the photometric orientation of
the morphological major axis is about right.  The major axis of the gas in
the model is also close to that observed for the two galaxies.

Item 13 notes that the bridge and tail of IC~2163 are nearly
symmetrical in HI. In the models, the bridge arm is more diffuse, but
still extensive.

Item 14 notes $100\ $km s$^{-1}$ streaming motions in the IC~2163
tail. A similar velocity spread is found in the model tail:
see Fig. 9. Item 15 notes that the mean velocity is nearly
constant along the tail. It has a shallow gradient in the models as
well; see Fig. 9.

The stellar arm contrast is large in both models and observations
(item 16). 

The spiral arms evident inside the IC~2163 oval (item 17) are not
obvious in the model, but our model may not have sufficient angular
resolution for discerning these thin arms. In the models, pre-existing
flocculent arms and small scale shocks both apparently contribute to
producing the dust filaments (item 18) in the tail of IC~2163 (see
Sec. 4.1.2). \citet{elm95a} point out an S-shaped wiggle of HI
emission from the tidal tails, which might be the result of a small
scale shock.

The spiral arms in NGC~2207 (item 20) are also not produced in any of
our models. Given the prominence and extent of these arms in NGC~2207,
and their complete absence in models like Sm1 (Fig. 8), it appears
that they must have been present before the collision. Thus, they must
be input into the models.

Item 21 (see also 26) reminds us that the HI distribution in the
NGC~2207 disc is dominated by a large partial ring (see Fig. 5). To
reproduce this the initial condition for galaxy A requires a large gas
hole in its centre.  The observed ring is then the disturbed remnant
of the initial annular gas distribution.

NGC~2207 is embedded in a large elliptically-shaped pool of HI gas
that extends a considerable distance to the southeast of the main disc
(item 22, Fig. 2). The model A disc is similarly distorted, but to a
lesser degree (Figs. 5, 8). Much of the southeastern extension in the
fiducial model is due to the spiral arm extension there. The adopted
gravitational potential may produce a force that is too strong in the
outer disc, preventing more gas from moving out there.

Inspection of Fig. 7 shows that the photometric PA (item 23) of the
model is close to that observed at a slightly earlier time if we assume
that both are dominated by the spiral arm morphology. Table 3 shows
that the gas kinematic PA in the model is also a reasonable fit to the
observation.

The radio continuum ridge in the NE quadrant of NGC~2207 (see
\citealt{elm95a}) is recalled in item 25. The fiducial model does not
exactly reproduce this feature. At earlier times the interaction
between the bridge and the outer spiral arm in the north is very
strong and would produce a ``continuum ridge'' (e.g., see panel 4 of
Fig. 5). Before and at the present time there is a similar, but
weaker, interaction between the bridge particles and the middle NE
spiral arm, which could be responsible.

This completes our review of Table 1, and we turn to the smaller-scale
features of Table 2, which begins with the luminosity and size
distributions of young star clusters in the system (item 27). The
study of star cluster structure is beyond the scope of our present
models. Item 29 notes the widespread distribution of star clusters in
the system. This is also seen in the fiducial model, see Sec. 6. The
model suggests an explanation in terms of widespread distribution of
compressed regions, especially in spiral arms and tidal structures.

Item 28 describes massive gas clumps without young star clusters. It
is true that such clumps can also be found in the spirals of Galaxy A
and the tail of Galaxy B. However, comparing the model and
observational clumps is problematic. The size of the observed clumps
is close to the effective beam size, so they are not
well-resolved. The scale of the model clumps is close to that over
which local self-gravity is computed in the disc. Moreover, they do
not contain a large number of particles, so they too are not well
resolved. On the other hand, the structure of the model clumps is more
reminiscent of knotty dust filaments seen in the spiral arms overlying
IC~2163 (item 32).

The lack of nuclear activity in this system is noted in items 30 and
34. This is replicated, at the present time, in the model SF. The
models suggest strong central compressions and enhanced SF in the core
of IC~2163 at earlier times (Sec. 6). They suggest that no large
compressions have occurred in the primary, and that most of the mass
transfer was initially into the outer disc.

The dust spirals in the center of NGC~2207 (item 31) are not seen in
the models. However, the models do suggest the passage of a number of
complex waves over the course of the encounter, and these would be a
plausible cause of the dust compression, along with some mass
transfer. 

Finally, item 33 describes the luminous SF region on the far western
arm of NGC~2207. This intense emission source could be an effect of an
earlier mass transfer interaction. However, we can only speculate
about this; it is extremely difficult to produce such small-scale
features in models at the present time. 

We should emphasize that the models are generally very successful at
reproducing the large-scale collisional morphology and kinematics, as
well as the qualitative SF history. The shortcomings of the models can
be grouped into several categories. The first category includes
structures that are well beyond the resolutions of current models,
like the young cluster luminosity functions, or very-small-scale
structures like item 33. The latter example falls in a subcategory
mentioned at the beginning of this section, features that might be
produced rarely in a model grid, but essentially by accident. The
second category includes quantities that can be very time-dependent in
the models, like the SF ratio of the two galaxies. A third category
includes features that are prominent, but do not seem to appear in any
of the models of a grid that is otherwise generally successful at
producing the observed collisional structure. We have suggested that
the spiral arms in these two galaxies, and possibly the warp of
NGC~2207, may fall in this category. They apparently predate the
collision and must be added to the initial conditions by hand.

The final category, which includes most of the discrepancies cited
above, is the set of things that appear 'fixable' if we tune the model
finely enough. In the present case, many of these relate to effects of
the detailed structure of the rigid gravitational potentials adopted,
and also to the dynamical friction treatment. There are two major
difficulties in making such fixes: 1) the precise change needed is not
clear {\em a priori}, and 2) such changes generally affect other structures
in a way that requires additional fixes. This can lead to a
cascade of 'fixes' that does not readily converge.

\section{Conclusions}

The discussion of the previous section illustrates the mounting
complexity involved in producing successful models of finer scale
features or more dynamically evolved systems. To significantly refine
some of the category 4 defects, while maintaining good fits to other
structures, would take considerably more runs. This is a good
procedure for the improvement of early runs where the experienced
modeler can fairly quickly identify the source of major defects
resulting from particular initial orbital or structural
parameters. After several dozen simulations have been run the
resulting models are usually quite good, but further improvements must
overcome the problems cited in the previous paragraph.

An automated procedure for producing and selecting the best of
hundreds of models would be desireable (e.g., like the N-body, genetic
algorithmic modeling of M51 by \citealt{wah01}). However, a procedure
that selects initial conditions that are distributed evenly in
parameter space is unlikely to be very efficient, since there are many
parameters, and the set of model initial conditions that yield a
sufficiently accurate result will generally have a very small volume
in parameter space. At the least, the parameter selection algorithm
will have to be guided in a nonlinear way by the results of previous,
and perhaps, some expert ``rules of thumb.'' To date, attempts at
automated modeling have been very limited.

On the positive side, it is not clear that modeling of specific
systems needs to be much more accurate than the best, current, by-hand
efforts. To reconstruct the complete dynamical history of an evolved
merger would require a great deal more, but there is not yet much
motivation for such an exercise. The present case shows that by-hand
modeling, with sequential improvements, can produce a result that
accounts for most of the morphology and kinematics in a pre-merger
collisional system. The present case and other detailed studies of
relatively symmetric pre-merger systems indicate that most of the
collision-induced SF can also be accounted for by such models (see
references in the introduction). Thus, we can use such systems as
laboratories for the study of some types of induced SF, while other
types, such as those found in merger remnants, require statistical
studies of samples of galaxies and models.

Finally, we note one unexpected result of the present modeling effort
is an indication that NGC 2207 has a rather smaller halo than the
Milky Way. Whereas the Milky Way halo may extend to 100 kpc or more
\citep{ive04}, our models suggest that the halo density of NGC
2207 may begin to tail off at roughly its optical isophotal
radius. Models of binary encounters between galaxies such as the one
presented here may eventually be useful for constraining the extent of
dark matter halos (see also \citealt{sal00a})

\section*{Acknowledgments}
We are grateful to the referee for his detailed examination of this
paper and many helpful comments. 

%Thanks to 

\bsp

\label{lastpage}

\end{document}